\documentclass{sig-alternate-05-2015}

\usepackage{mahnigCEC}
\usepackage{xcolor,colortbl}
\usepackage{epsfig}
\usepackage[hidelinks]{hyperref}
\usepackage{multirow}
\usepackage{array}
\usepackage[latin1]{inputenc}
\usepackage[T1]{fontenc}
\usepackage{bigstrut}
\usepackage{url}
\usepackage{cite}
\usepackage{booktabs}
\usepackage[pass,letterpaper]{geometry}
\begin{document}

\setcopyright{acmcopyright}

%

%

%

\CopyrightYear{2016}
\setcopyright{acmcopyright}
\conferenceinfo{GECCO '16,}{July 20-24, 2016, Denver, CO, USA}
\isbn{978-1-4503-4206-3/16/07}\acmPrice{\$15.00}
\doi{http://dx.doi.org/10.1145/2908812.2908914}

\title{Evolutionary Approaches to Optimization Problems in Chimera Topologies}

\numberofauthors{4}
\author{
\alignauthor 
 Roberto Santana\\
      \affaddr{University of the Basque Country (UPV/EHU)}\\
       \affaddr{San Sebasti\'{a}n, Spain}\\
       \email{roberto.santana@ehu.es}
\alignauthor
 Zheng Zhu\\
     \affaddr{Texas A\&M University}\\
    \affaddr{College Station, Texas, US}\\
   \email{zzwtgts@tamu.edu}     
\alignauthor 
    Helmut Katzgraber\\
      \affaddr{Texas A\&M University}\\
     \affaddr{College Station, Texas, US}\\
    \email{katzgraber@physics.tamu.edu}
}

\maketitle
\begin{abstract}
  Chimera graphs define the topology of one of the first commercially available quantum computers. A variety of optimization problems have been mapped to this topology to evaluate the behavior of quantum enhanced optimization heuristics in relation to other optimizers, being able to efficiently solve problems classically to use them as benchmarks for quantum machines. In this paper we investigate for the first time the use of Evolutionary Algorithms (EAs) on Ising spin glass instances defined on the Chimera topology. Three genetic algorithms (GAs) and three estimation of distribution algorithms (EDAs) are evaluated over $1000$ hard instances of the Ising spin glass constructed from Sidon sets. We focus on determining whether the information about the topology of the graph can be used to improve the results of EAs and on identifying the characteristics of the Ising instances that influence the success rate of GAs and EDAs. 
\end{abstract}



\section{Introduction}

Quantum computation (QC) aims to profit from the quantum properties of elementary particles to devise new, more efficient ways for representing and manipulating information. A number of paradigms have been proposed as promising candidates for QC. One of these paradigms is adiabatic quantum computing \cite{Santoro_et_al:2002,Brooke_et_al:1999} and perhaps the best known implementation of this approach for practical computation is the D-Wave quantum machine \cite{Dahl:2013}.

  In very simple terms, the D-Wave architecture consists of an array of qubits  coupled following a predefined topology (Chimera graph). Qubits have  associated weights and each coupler has an associated strength. Current D-Wave quantum annealing processors are designed to minimize the energy of an Ising spin configuration whose pairwise interactions lie on the edges of a Chimera graph.  D-Wave computers have served as an excellent testbed for investigating whether QC is actually possible. In particular, many studies have focused on the behavior of the quantum annealing (QA) optimization method  \cite{Kadowaki_and_Nishimori:1998,Chakrabarti_et_al:2005}.  QA is similar to simulated annealing (SA) but transitions between states (solutions) do not depend on thermodynamic fluctuations. These transitions are determined by quantum fluctuations. 

Although other optimization problems of arbitrary pairwise interaction structure and manageable size can be embedded into the D-Wave topology, a more straightforward approach to study the performance of this QC paradigm is to address problems whose structure  has been defined on the Chimera graph. Understanding which are the characteristics of problems defined on Chimera graphs is thus a very important issue. The behavior of a number of algorithms have been investigated  for problems with Chimera graph structure. On these problems, QA has been compared to  optimization methods such as Tabu search \cite{Mcgeoch_and_Wang:2013}, SA \cite{Katzgraber_et_al:2015} and parallel tempering (PT) \cite{Zhu_et_al:2015}. 

In this paper we propose a through investigation of evolutionary algorithms (EAs) for Ising spin glass problems defined on Chimera graphs. Our goal is to determine to what extent the search principles in which EAs are based on can be valuable for the solution of these problems. We argue that EAs can and should be used not only as solvers but also as a tool to better understand the characteristics of relevant problems. In particular, identifying sources of problem difficulty that are common to optimization approaches with completely different search strategies can advance the understanding of the problems and of the optimizers. We show how EAs can be used with these goals in mind and provide evidence of the benefits of this approach. 

  To study the target problems, we apply hybrid genetic algorithms (GAs) and variants of estimation of distribution algorithms (EDAs) based on factorizations \cite{Baluja_and_Davies:1997r,Muhlenbein_et_al:1999,Shakya_and_Santana:2012}. This strategy allows us to investigate EAs that are blind to the problem structure and other variants that exploit the information about this structure for a more efficient search. In addition to elucidate which is the difficulty that Chimera-graph based spin glass instances pose to EAs, we expect to obtain insights from the behavior of EAs on these instances.  To the knowledge of the authors problems defined on Chimera graphs have not been previously addressed using EAs. 

The paper is organized as follows: The next section presents  some necessary background on Chimera graphs and the Ising model.  Section~\ref{sec:RELWORK} discusses related work on the relevant questions treated in the paper. Section~\ref{sec:EAs} presents the EAs used to optimize the instances. Section~\ref{sec:EXPE} introduces the experimental framework used to investigate our hypothesis,  presents the numerical results and discusses our findings. Section~\ref{sec:CONCLU} concludes the paper.

\section{Background} \label{sec:BACKGROUND}

\subsection{Ising model}

A Hamiltonian function serves to describe how the energy of a physical process depends on the state of the system's particles and their interactions.  The generalized Ising model  is described by the Hamiltonian shown in Equation~\eqref{ISINGM} where $L$ is the set of sites called a lattice. Each spin variable $\sigma_i$ at site $i \in L$  either takes the value $1$ or $-1$. One specific choice of values for the  spin variables is called a configuration. The constants $J_{ij}$  are the  interaction coefficients or couplings. The ground state is the configuration with minimum energy.  
  
 \begin{equation}  
 H = - \sum_{<i,j> \in L} J_{ij} \sigma_i \sigma_j - \sum_{i \in L} h_{i}  
 \sigma_i  \label{ISINGM}  
 \end{equation}  
where the first sum is over pairs of spins adjacent in the lattice $L$.
  
Spin couplings can take values from an a priori defined set of values and both the choice of the set and the distribution with which the couplings are sampled from it influence the characteristics of the problem and its difficulty for the optimizers. In this paper we use couplings selected from Sidon sets of the type $U_{5,6,7} = \{\pm 5,\pm 6, \pm 7\}$ as advocated in  \cite{Katzgraber_et_al:2015}. The rationale behind this choice of the couplings is explained in Section~\ref{sec:EXPBENCHMARK}.  We consider instances for which $h_{i}=0, \;\forall{i \in  L}$.

\subsection{D-Wave architecture and Chimera graphs}

\begin{figure}[htb] 
\begin{center} 
\begin{pspicture}(0,0)(6.0,6.0) 

\rput(0.75,5){\Circlenode[radius=0.3cm]{P0}{$ 0$}} 
\rput(2.5,5){\Circlenode[radius=0.3cm]{P1}{$ 4$}} 
\rput(0.75,3.5){\Circlenode[radius=0.3cm]{P2}{$ 1$}}
\rput(2.5,3.5){\Circlenode[radius=0.3cm]{P3}{$ 5$}}
\rput(0.75,2){\Circlenode[radius=0.3cm]{P4}{$ 2$}}
\rput(2.5,2){\Circlenode[radius=0.3cm]{P5}{$ 6$}} 
\rput(0.75,0.5){\Circlenode[radius=0.3cm]{P6}{$3$}} 
\rput(2.5,0.5){\Circlenode[radius=0.3cm]{P7}{$7$}}

\ncline{-}{P0}{P1} \ncline{-}{P0}{P3}  \ncline{-}{P0}{P5}  \ncline{-}{P0}{P7} 
\ncline{-}{P1}{P2} \ncline{-}{P1}{P4}  \ncline{-}{P1}{P6}
\ncline{-}{P2}{P3} \ncline{-}{P2}{P5}  \ncline{-}{P2}{P7}  
\ncline{-}{P3}{P4} \ncline{-}{P3}{P6} 
\ncline{-}{P4}{P5} \ncline{-}{P4}{P7} 
\ncline{-}{P5}{P6}  
\ncline{-}{P6}{P7}

\rput(3.75,5){\Circlenode[radius=0.3cm]{P8}{$8$}} 
\rput(5.5,5){\Circlenode[radius=0.3cm]{P9}{$12$}} 
\rput(3.75,3.5){\Circlenode[radius=0.3cm]{P10}{$9$}}
\rput(5.5,3.5){\Circlenode[radius=0.3cm]{P11}{$13$}}
\rput(3.75,2){\Circlenode[radius=0.3cm]{P12}{$10$}}
\rput(5.5,2){\Circlenode[radius=0.3cm]{P13}{$14$}} 
\rput(3.75,0.5){\Circlenode[radius=0.3cm]{P14}{$11$}} 
\rput(5.5,0.5){\Circlenode[radius=0.3cm]{P15}{$15$}}

\ncline{-}{P8}{P9} \ncline{-}{P8}{P11}  \ncline{-}{P8}{P13}  \ncline{-}{P8}{P15} 
\ncline{-}{P9}{P10} \ncline{-}{P9}{P12}  \ncline{-}{P9}{P14}
\ncline{-}{P10}{P11} \ncline{-}{P10}{P13}  \ncline{-}{P10}{P15}  
\ncline{-}{P11}{P12} \ncline{-}{P11}{P14} 
\ncline{-}{P12}{P13} \ncline{-}{P12}{P15} 
\ncline{-}{P13}{P14}  
\ncline{-}{P14}{P15}

\ncarc[arcangle=35]{-}{P0}{P9}
\ncarc[arcangle=35]{-}{P2}{P11}
\ncarc[arcangle=35]{-}{P4}{P13}
\ncarc[arcangle=35]{-}{P6}{P15}
\end{pspicture} 
 \caption{Detail of a Chimera graph. Two connected $K_{4,4}$ blocks.}
 \label{fig:Knn}
\end{center} 
\end{figure}
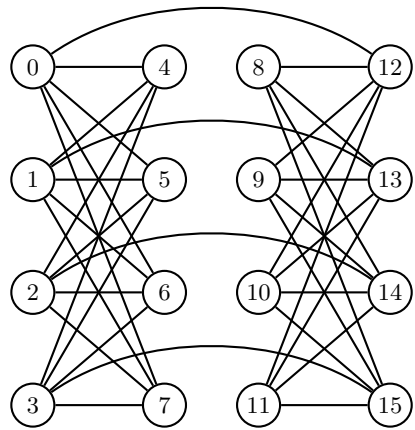

The D-Wave architecture follows a regular repeating pattern to tile out a processor \cite{Dahl:2013}.  The building blocks of the Chimera graph are symmetric subgraphs referred to as $K_{4,4}$ graphs.  The eight qubits included in the $K_{4,4}$ graph are connected as bipartite graphs. One example of  two connected $K_{4,4}$ graphs  is shown in Figure~\ref{fig:Knn}.  In addition to bipartite connections, in each group, the four left nodes are also connected to their respective north/south grid neighbors and the $4$ right nodes are connected to their east/west neighbors. Thus internal nodes have degree $6$ and boundary nodes have degree $5$. The Chimera graph is formed by $n^2$ $K_{4,4}$ subgraphs.

\section{Related work}  \label{sec:RELWORK}

The  questions treated in this paper are related to different lines of research in the fields EAs and also to studies on the behavior of different types of optimizers for Chimera graphs. Therefore, we organize the review of previous work according to these relevant topics. 

\subsection{Previous EA approaches to Ising and other problems from Physics}

   Ising spin glasses are not only relevant because of the role they play in  Physics. They represent a formidable problem to investigate the capabilities of different EAs and as such they have been investigated in a number of papers. 

  GAs were early applied to Ising models \cite{Anderson_et_al:1991,Maksymowicz_et_al:1994,Prugel_and_Shapiro:1997}. Some of these works used knowledge about the topology of the problem to design the genetic operators.  In \cite{Anderson_et_al:1991}, a block-crossover, able to exploit the 2-dimensional structure of the instances was proposed. 
     EDAs have been also used to solve Ising models \cite{Pelikan_and_Hartmann:2006,Santana:2005,Shakya_et_al:2006}. In \cite{Pelikan_and_Hartmann:2006}, the Bayesian optimization algorithm (BOA) was applied to instances with $\{\pm 1\}$ and Gaussian couplings. A deterministic hill climber based on single-bit flips was used to solve problems of up to $400$ spins. To solve bigger instances of $512$ and $1024$ spins, a more sophisticated cluster exact approximation method was required. In general, successful EDAs applications to large Ising models are in fact hybrid algorithms that incorporate local search methods.

While Ising problems addressed with EAs have been generally defined on regular 2-d and 3-d grids, other topologies have been also used.   In \cite{Pelikan_and_Katzgraber:2009}, the behavior of EAs was investigated on one-dimensional spin glass with power-law interactions.  In \cite{Echegoyen_et_al:2010}, the impact of different network topologies (e.g.  grids, small-world networks and random graphs) of the underlying problem  on the models learned by EDAs was investigated. We did not find  previous reports on the application of  EAs to Ising models defined on Chimera graphs. It is important to notice that the Chimera topology considerably  departs from regular grids.

\subsection{Exploiting the problem structure}
 
 Using problem information to improve EAs has been always a relevant issue in evolutionary computation (EC).  One relevant question is whether and how  knowledge about the underlying topology of the optimization problem can be used to increase the efficiency of the EA approaches to this problem. In this paper, we investigate this issue in three different ways: 1) Using the Chimera structure to design building-block  wise crossover operators where the building blocks corresponds to groups of variables in the same $K_{4,4}$ subgraph. 2) Using factorized EDAs in which the factors comprise variables in each $K_{4,4}$ subgraph. 3) Biasing the search for tree-based EDAs to include dependencies between variables that are connected in the Chimera structure. 

Building-block wise crossover is at the core of successful probabilistic model building GAs like the extended compact GA \cite{Harik_et_al:2006}. EDAs that explicitly use to different extent the regular grid structure of the Ising problems have been previously proposed in \cite{Santana:2005,Shakya_et_al:2006}. EDAs that bias the learning of probabilistic trees using a priori information about the problem has shown to be more efficient \cite{Baluja:2006}.

 Other methods that use information about the structure of the graphs to solve Ising problems defined on Chimera graphs have been also proposed in the field of Physics. In \cite{Zintchenko_et_al:2015}, a heuristic method that finds optimal configurations of local clusters of spins as a way to reach the ground state is presented. The target clusters are those that are strongly coupled to each other and more weakly coupled to the rest of the system. Although the factorized EDA we use in our comparisons use information about clusters of solutions, these clusters depend on the Chimera graph and not on the strength of the couplings. 

The Hamze-de Freitas-Selby (HFS) algorithm \cite{Hamze_and_deFreitas:2004,Selby:2014}, uses  a subgraph-based sampling method for Ising-type models with frustration. A collection of subgraphs induced by the original topology are used instead of single spins for a more efficient sampling with GS and PT. In Selby \cite{Selby:2014}, experiments on Chimera graphs were done using trees as the induced subgraphs. In this paper, we use EDAs based on tree models. However, the trees are learned from the selected solutions by applying statistical methods.

\subsection{Fitness landscape analysis and investigation of fitness difficulty} 

 A number of papers \cite{Dash:2013,Mcgeoch_and_Wang:2013} have investigated the behavior of classical optimizers on Ising instances defined on Chimera graphs.  In  \cite{Mcgeoch_and_Wang:2013},  three conventional software solvers: CPLEX, a variant of Tabu, and a branch-and-bound search, were compared to QA in Chimera-structured problems. 

 In \cite{Isakov_et_al:2015}, different variants of SA were evaluated on Ising problems defined on Chimera graphs. However, the focus of the paper was not on the comparison of SA with other algorithms but on the development of fast SA implementations. In \cite{Katzgraber_et_al:2015}, an extensive comparison between SA and QA is presented for Ising problems with Chimera topology. This work  identified the important effect that the choice of the couplings for the Ising instances could have in the comparison between optimizers for problems defined on Chimera-graphs. Instances generated using Sidon sets\footnote{In a Sidon set, the sum of two members of the set gives a number that is not part of the set} were proposed as harder to optimize.  Also in \cite{Zhu_et_al:2015}, a particular class of Sidon instances ($J_{ij} = \{5,6,7\}$)  was used as a test-bed for evaluating the behavior of QA which was compared to PT.

\section{Evolutionary optimization approaches} \label{sec:EAs}

 Let ${\bf{X}}=(X_1,\ldots ,X_n)$ denote a vector of discrete random variables where $n$ is the number of nodes of the Chimera graph. We use ${\bf{x}}=(x_1,\ldots,x_n)$ to denote an assignment to the variables.  In our problem representation, each binary variable represents one spin configuration of a node in the Chimera graph.  In the chromosome representation, variables are ordered according to the position of the nodes in the Chimera graph. For instance, the first $16$ variables would correspond to the nodes shown in Figure~\ref{fig:Knn}. Therefore, the variables that represent nodes that are connected in the Chimera graph are, in most of the cases, relatively close in the chromosome.  The fitness function used to evaluate the solutions is the one represented by Equation~\eqref{ISINGM}.

\subsection{Benchmarked algorithms}

Algorithm~\ref{alg:EA} shows  the general pseudocode of all the EAs tried in the paper. Population size $N=512$ and truncation selection with parameter $T=0.5$ were used.

\begin{BAlgo}{EA} 
 \label{alg:EA} 
 \item Set $t\Leftarrow 0$. Generate an initial population $D_0$ of $N\gg 0$ random solutions.  
 \item \Do 
 \item \T {For each solution, apply a greedy local search method and output local optimum.}
 \item \T {Select from population $D_t$ a set $D_t^S$ of $k \leq N$ points using truncation selection.}
 \item \T {Generate a new population $D_{t+1}$ from $D_t^S$ applying the variator operator of choice.}
 \item \T {Apply random bit-flip mutation to solutions in $D_{t+1}$.}
 \item \T {$t \Leftarrow t+1$}  
 \item  \Until{Termination criteria are met.}  
\end{BAlgo}

  In the design of EAs, certain assumptions about the structure of the optimization problem are usually implicitly or explicitly made. The extent to which these assumptions agree with the characteristics of the problem usually determine the success of the EA. We have selected six EAs by considering how their mechanisms to explore the space of solutions may be related with the structure of the Chimera graphs.

\begin{enumerate}

 \item GA with 1-point crossover (1PCX-GA): Simple GA algorithm with a crossover probability of $1$ and where a crossover point is randomly selected between positions $2$ and $n-1$. The two offspring are created by taking one segment from each parent.   
 \item GA with uniform CX (uCX-GA): Similar to 1PCX-GA but the alleles of the offspring are randomly taken from each of the two parents with probability $0.5$.

 \item GA with bit-wise CX (BWCX-GA): Similar to uCX-GA but instead of single bits, blocks of variables are taken from each parent. These blocks correspond to variables that are related by the Chimera graph. In particular, there is one block for each subgraph $k_{4,4}$ involving $8$ variables and for each edge joining these subgraphs.

 \item Factorized distribution algorithm (FDA): An EDA that uses the same structure as BWCX-GA. Marginal probabilities for all configurations of the blocks are learned from the selected solutions and new solutions are generated sampling from a junction tree \cite{Muhlenbein_et_al:1999} constructed from this factorization.

 \item Tree-based EDA (Tree-EDA): An EDA that learns the structure of the graphical model from the matrix of mutual information between \emph{all possible pairs} of variables \cite{Baluja_and_Davies:1997r}. The pairs with the strongest mutual information are used to build a tree structure from which solutions are sampled. 
 
 \item Tree-based EDA with a priori information (Tree-EDA$^r$): Idem to Tree-EDA but only pairs of variables connected in the Chimera topology are used to learn the probabilistic model. 
\end{enumerate}

\subsection{Efficient local search}

 In step $3$  of Algorithm~\ref{alg:EA}  a greedy local search method was added as part of the evaluation process.  Previous works \cite{Pelikan_and_Hartmann:2006,Santana:2005} have shown that without the inclusion  of local optimizers,  EAs are not expected to be competitive algorithms for large Ising spin glass problems. 

 The details of the greedy local  search implemented are shown in Algorithm~\ref{alg:GSEACH}. One main characteristic of the method is that the local fitness value of the current solution are stored in memory and only local computations are needed to evaluate the effect of bit flips. Therefore, only a fraction of computations are needed to evaluate the $n$ possible bit-flips possible from the current solution.  In addition, when the initial solution is a local optimum and no bit-flip improves the current fitness value, the local optimizer allows a predefined number of transitions ($5$) to solutions with lower fitness. In these cases, the bit-flip that decreases the fitness the least is selected. This mechanism, similar to one of the components of Tabu search  was designed as a way to help the algorithm to escape from local optima.  However, the local optimizer does not implement a memory or any other type of more sophisticated components. The local optimizer stops when the current solution can not be improved or the maximum number of transitions to solutions of lower fitness has been consumed. 

 \begin{BAlgo}{Greedy search}
 \label{alg:GSEACH}
 \item  Set current solution to initial solution. 
 \item  Initialize number of accepted negative $n_{neg\_moves}=0$.
 \item  Evaluate the $2$ possible assignments for the $n$ variables of the initial solution keeping other variables fixed.
 \item \Do
 \item \T{Select the variable $X_i$ whose bit-flip improves the fitness the most}
 \item \T{If improvement is negative, initial solution was already improved or $n_{neg\_moves}=max_{neg\_moves}$}
 \item \TT{Return current best solution}
 \item \T{Else if improvement is negative, initial solution has not been improved, and $n_{neg\_moves}<max_{neg\_moves}$.}
 \item  \TT{$n_{neg\_moves} \rightarrow  n_{neg\_moves}+1$}
 \item  \T{Bit-flip $X_i$ in current solution.} 
 \item  \T{Update local fitness values for variable $X_i$ and all its neighbors in the Chimera graph.} 
\item   \While{}
\end{BAlgo}

After solutions have been evaluated a  random bit-flip mutation operator was applied. Extensive preliminary experiments showed that this way of infusing diversity in the population was a required ingredient for avoiding early convergence to poor solutions. More remarkably, a high mutation rate of $p_m=0.2$ showed to produce the best results. Notice, that all the EAs are compared in the same conditions. The only difference between the implementation of the six EAs is in step $5$ of Algorithm~\ref{alg:EA}. This step comprises the recombination operator for the GAs and the learning and sampling step for EDAs. The termination criteria for all algorithms was a maximum of $1000$ generations or reaching a low diversity in the population ($10$ or less genotypically different individuals). 

\subsection{Structural hypothesis and EAs}  

 The choice of the EAs has been made with the aim of evaluating different questions about the best EA approach to problems with a Chimera structure. We briefly state these questions and their  relation to the algorithms.

 \begin{itemize}
  \item How relevant is the impact of potential building block disruption? 1PCX-GA and uCX-GA differ only in the type of crossover mechanism they use. Uniform crossover is a more disruptive operator.  By evaluating the difference in the performance of these two algorithms we can have a rough idea of the importance of respecting the interactions of the problem.

  \item Which is the added gain of considering the underlying structure of the Chimera graph in the design of the crossover operator? By comparing BWCX-GA to 1PCX-GA we can measure if a more informed choice of the information to exchange between the parents has a strong effect in the performance of the GAs.
  \item Can structured-informed factorized approximations of the selected solutions make a difference over ``blind''  and more intelligent crossover operators?  By comparing the FDA, which uses the blocks of variables related in the Chimera graph as factors,  to all other GAs we can determine if using problem information the way EDAs do makes a difference for this problem.
  \item Should EDAs learn the structure of the interactions from the data instead of inferring them from the Chimera structure?  Are pair-wise interactions sufficient to solve the problem? By comparing Tree-EDA to FDA and GAs we can find answers to these questions. 
  \item Are the pair-wise interactions determined by the Chimera graph sufficient to solve the problem? The comparison between the performance of Tree-EDA and Tree-EDA$^r$ will help to answer this question.
 \end{itemize}

\section{Experiments} \label{sec:EXPE}

  The goal of the experiments is twofold. 1) We want to evaluate the difference in the behavior of the evolutionary algorithms and whether these differences offer clues about the characteristics of the Chimera instances. 2) We would like to identify which characteristics of the instances (descriptors) have an impact in the performance of EAs.

\begin{figure}[hbtp]
\begin{center}
 {\includegraphics[width=8.0cm]{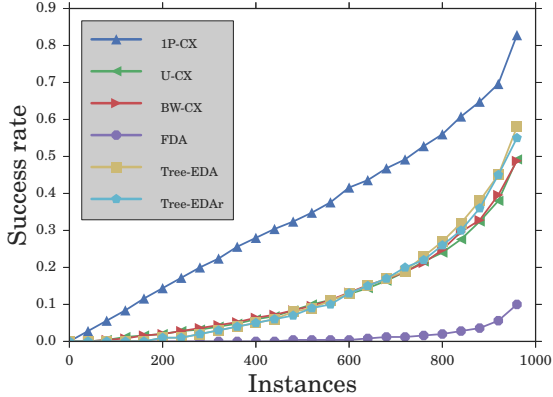}}
\caption{EAs success rate for the $1000$ instances considered. For each algorithm, instances have been sorted from lowest to highest success rate.} 
\label{fig:SUCCRATE}
\end{center}
\end{figure}

\subsection{Experimental benchmark} \label{sec:EXPBENCHMARK}

 To evaluate the behavior of the algorithms, $1000$ Ising instances defined  on the Chimera graph were used. These instances were proposed in \cite{Katzgraber_et_al:2015,Zhu_et_al:2015} and investigated to evaluate the behavior of QA in the D-Wave quantum annealer. Instances were generated with couplings defined in the $U_{5,6,7} = \{\pm 5, \pm 6, \pm 7\}$ Sidon set. 

 There are a number of reasons why the Sidon set instances have been particularly useful for studies in Physics \cite{Katzgraber_et_al:2015,Zhu_et_al:2015,Zhu_et_al:2015a} and relevant for us: 1) All Instances have unique (two if spin reversal symmetry is considered) ground states and all spins have nonzero local fields (by carefully selecting the combination of bonds), these features make instances harder than random bi-modal instances. 2)  With Sidon interactions, there is full control of the number of ground states and low lying excited sates, even energy gaps, so it is possible to study resilience of instances with more precision. 3)  With full control of many parameters of the instances, it would be possible to identify the important ones that are related to the behavior of different algorithms.
  
 For these instances, the ground states and first exited states with equal probabilities  have been found by applying the isoenergetic cluster algorithm \cite{Zhu_et_al:2015a} and running millions of Monte Carlo Sweeps for each instance. Recent results \cite{Mandra_et_al:2016} show that the isoenergetic cluster algorithm is one of the few algorithms that scales better than quantum annealing on Google instances \cite{Denchev_et_al:2015}. 

We use a number of descriptors that characterize the instances. These descriptors are derived from an analysis of the landscape of the spinglass order parameter distribution $P(q)$ for the instances. $P(q)$ overlap distributions is the proxy to the complexity of energy landscape of an instance.  It measures the  average Hamming distance between two solutions randomly sampled from the problem's low-temperature Boltzmann distribution.  This approach can be used to identify the instances with tall and thin energy barriers \cite{Mandra_et_al:2016}.  We considered the following descriptors:

\begin{enumerate}
  \item  GRstate: Fitness value of the ground state solution.  
  \item   dJ+dh\_RES:  Resilience of instance to coupler and qubit noise (upper bound success probability). Quantifies the robustness of ground-state configurations to noise in the D-Wave device. 
  \item    PEAK\_1, PEAK\_2:  Position of the highest and second highest peaks in the $P(q)$ distribution plot. 
  \item    HEIGHT\_1,HEIGHT\_2: Height of the highest and second highest peaks in the $P(q)$ distribution plot. 
  \item     H\_SCORE:  Ratio between the height of second highest peak and the  height of highest peak. 
  \item   NOISE\_SCR:  Ratio between the  height of third highest peak and the height of second highest peak.
\end{enumerate}

$250$ runs were executed for all the algorithms except for Tree-EDA and Tree-EDA$^r$ for which, due to the high computational time spent by the algorithms, only $100$ runs were executed.

\subsection{Behavior of EAs}

Table~\ref{tab:SUCCRATE} shows the number of instances for which the ground state was found in $r$ percentage of the runs. For  $r=0.25$, the table shows the number of instances whose optima were found at least once in all the runs. It can be seen that the best results were achieved by GAs over EDAs, with the best absolute results achieved by 1PCX-GA.  These results are better detailed in Figure~\ref{fig:SUCCRATE} where instances are sorted according to the success rate reached by each algorithm.  BWCX-GA and uCX-GA have a similar behavior, while these algorithms are more efficient than Tree-EDA and Tree-EDA$^r$ for finding the ground state  $10\%$ of the time or less frequently, the EDAs find more instances with $25\%$ and higher success rate, a fact that can be also observed in Table~\ref{tab:SUCCRATE}. 

\begin{table}
\begin{center}
\caption{Number of instances (out of $1000$) for which the ground state was found in $r \%$ of the runs. For Tree-EDA and Tree-EDA$^r$, $100$ runs were executed. For all other algorithms, $250$.} 
\scriptsize
\begin{tabular}{r|r|r|r|r|r|r}
\toprule
 r (\%)& 1P-CX & U-CX & BW-CX & FDA & Tree & Tree-r \\ 
\midrule
           0.25&        991 &        935 &         940 &    543 &      821 &       832 \\
           1   &        982 &        881 &         878 &    322 &      821 &       832 \\
           10  &        868 &        480 &         471 &     45 &      474 &       462 \\
           25  &        645 &        188 &         199 &      4 &      218 &       212 \\
           50  &        274 &         37 &          38 &      1 &       65 &        63 \\
           75  &         54 &          6 &           4 &      0 &       12 &        13 \\
           90  &         13 &          1 &           0 &      0 &        3 &         2 \\
\bottomrule
\end{tabular}
\label{tab:SUCCRATE}
\end{center}
\end{table}

The success rate of the algorithms does not provide the whole picture about their behavior since some EAs may exhibit a high variability being able to achieve high quality or poor solutions depending on the instances. Therefore, for each instance, we applied a multiple comparison statistical test to look for significant differences between algorithms using the best solutions reached in $100$  runs (i.e. not the number of times that the optimum was found in these runs). The Kruskal Wallis test was applied first, and by applying a posthoc test we looked for statistical differences between each pair of algorithms. A Bonferroni correction was added to compensate for multiple comparisons. All tests used as pvalue $\alpha = 0.01$.

  Table~\ref{tab:POSTHOCTEST} summarizes the results of the pair-wise tests for all instances. In the table, cell $(r,c)$ indicates the number of instances for which algorithm in row $r$ was significantly better than algorithm in column $c$. For example, algorithm 1PCX-GA was significantly better than FDA for the $1000$ instances. From the analysis of Table~\ref{tab:POSTHOCTEST} it is clear that 1PCX-GA significantly outperforms all other algorithms. There are not significant differences between the pair of algorithms (uCX-GA,BWCX-GA)  and the pair (Tree-EDA,Tree-EDA$^r$). This seems to indicate that using information about the structure of the problem does not provide any advantage for the search, at least this is the case for GA with uniform crossover and Tree-EDA. Also, a conclusion from the analysis is that FDA is not a good choice for this problem.

\begin{table}
\begin{center}
\caption{Summary of the pair-wise statistical tests. Cell $(r,c)$ indicates the number of instances for which algorithm in row $r$ was significantly better than algorithm in column $c$.}
\scriptsize
\begin{tabular}{c|r|r|r|r|r|r}
\toprule
 Alg.  & 1P-CX & U-CX & BW-CX & FDA & Tree & Tree-r \\ 
\midrule
    1P-CX    &        - &           862&         858 &    1000 &      671 &     707 \\
    U-CX    &        0 &           -  &           0 &     973 &      33 &       31 \\
    BW-CX    &        0 &           0  &           - &     974 &      39 &       41 \\
    FDA      &        0 &           0  &           0 &       - &       0 &        0  \\
    Tree     &        2 &         201  &          196&     980 &       - &        0  \\
    Tree-r   &        2 &         182  &          173&     971 &       0  &        - \\           
\bottomrule
\end{tabular} 
\label{tab:POSTHOCTEST}
\end{center}
\end{table}

\begin{figure}[hbtp]
\begin{center}
{\includegraphics[width=8.0cm]{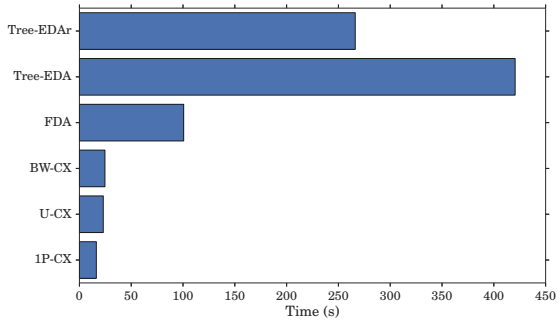}}
\caption{Average computational time of the algorithms in solving all the instances.}
\label{fig:COMPTIME}
\end{center}
\end{figure}

  Figure~\ref{fig:COMPTIME} shows the average computational time of all the algorithms across all instances. It can be appreciated that 1PCX-GA is also the fastest  among all the EAs compared. It is slightly faster than the other two GAs because uniform crossover requires the systematic generation of random numbers during the crossover and in 1PCX-GA only one random number has to be generated for each crossover. All EDAs require a higher computational time than any GA. In particular Tree-EDA is approximately $8$ times slower than 1PCX-GA. As expected, since Tree-EDAs need to learn the structure of the model from data, they are more computationally costly than FDA. Also, there is a clear gain in efficiency in Tree-EDA$^r$ over Tree-EDA. This gain is due to considering for the construction of the tree only the pairwise relationships that exist in the Chimera graph.

\subsection{Relationship with instance descriptors}

 Another important question is to determine whether and how is the performance of the EAs related to the characteristics of the instances. This issue is particularly relevant since the Sidon instances were originally engineered to investigate the behavior of QA on the D-Wave computers. Unveiling this type of relationship could help to find links between the behavior of EAs and optimizers that use completely different search mechanisms. In addition, by investigating this relationship we can determine whether the instance descriptors have a similar impact on all EAs or some of the descriptors are better signatures of behavior for some algorithms than for others.

\begin{table}
\begin{center}
\scriptsize
\caption{Correlation between the descriptors of the instance and success rate of the EAs. Correlations that were \emph{not} found significant for $p=0.01$ are underlined.} 
\begin{tabular}{l|r|r|r|r|r|r}
\toprule
 Descriptors & 1P-CX & U-CX & BW-CX & FDA & Tree & Tree-r \\ 
\midrule
     GRstate &      -0.21 &      -0.17 &       -0.17 &  -0.14 &    -0.17 &     -0.18 \\
   dJ+dh\_RES &       0.30 &       0.22 &        0.22 &   \underline{0.07} &     0.16 &      0.16 \\
      PEAK\_1 &       0.16 &       0.11 &        0.12 &   \underline{0.03} &     0.12 &      0.12 \\
    HEIGHT\_1 &       0.20 &       0.17 &        0.17 &   0.08 &     0.19 &      0.18 \\
     PEAK\_2 &       0.19 &       0.13 &        0.12 &  -0.00 &     \underline{0.07} &      \underline{0.07} \\
    HEIGHT\_2 &       \underline{0.05} &       \underline{0.02} &        \underline{0.02} &   0.00 &    \underline{-0.06} &     \underline{-0.04} \\
     H\_SCORE &      -0.15 &      -0.13 &       -0.13 &  \underline{-0.07} &    -0.17 &     -0.15 \\
   NOISE\_SCR &      -0.32 &      -0.27 &       -0.27 &  -0.09 &    -0.20 &     -0.19 \\  
\bottomrule
\end{tabular}
\label{tab:DES_SUCCRATE}
\end{center}
\end{table}

\begin{figure*}[hbtp]
\begin{center}
 \includegraphics[width=5.5cm]{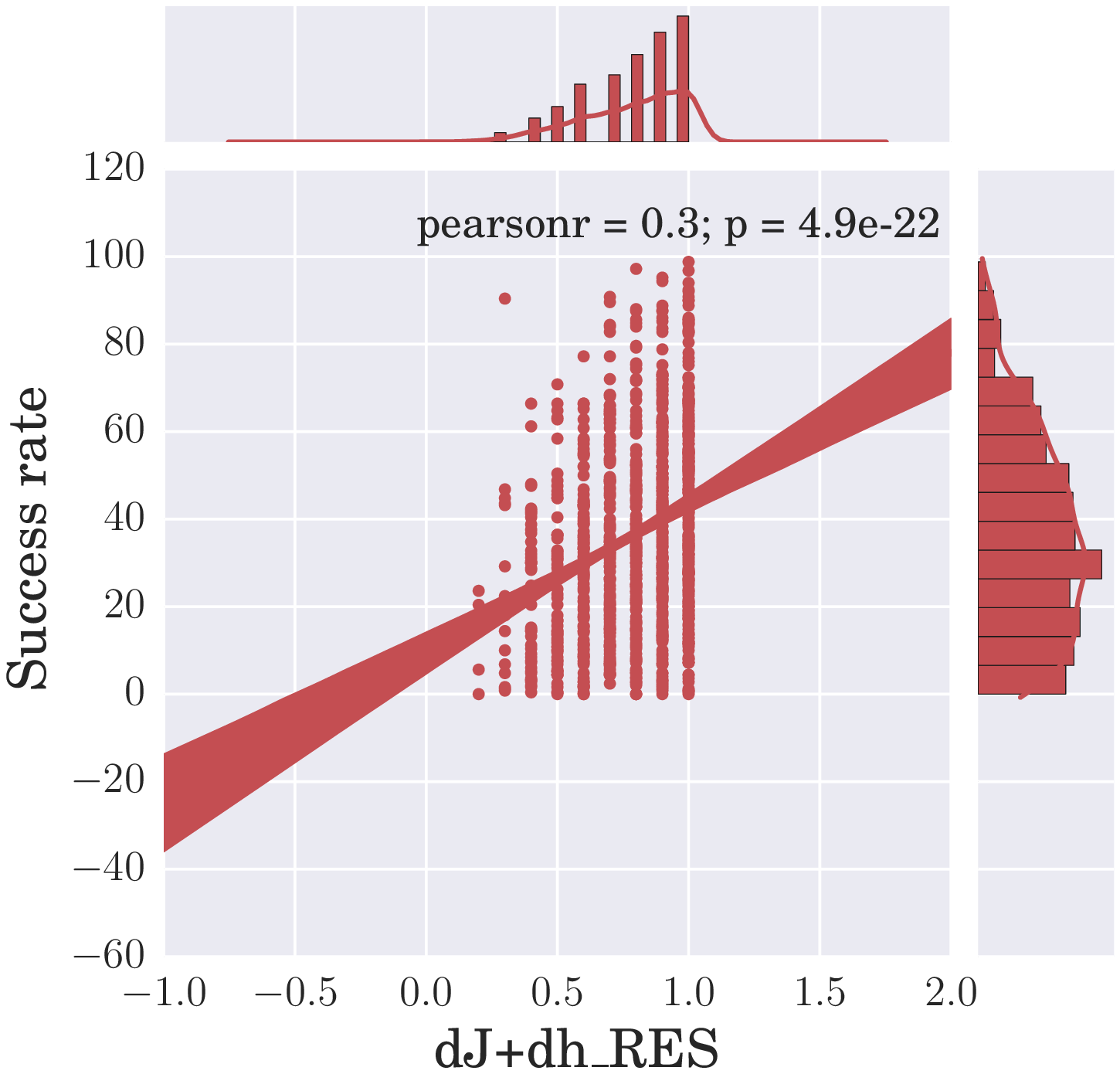}
 \includegraphics[width=5.5cm]{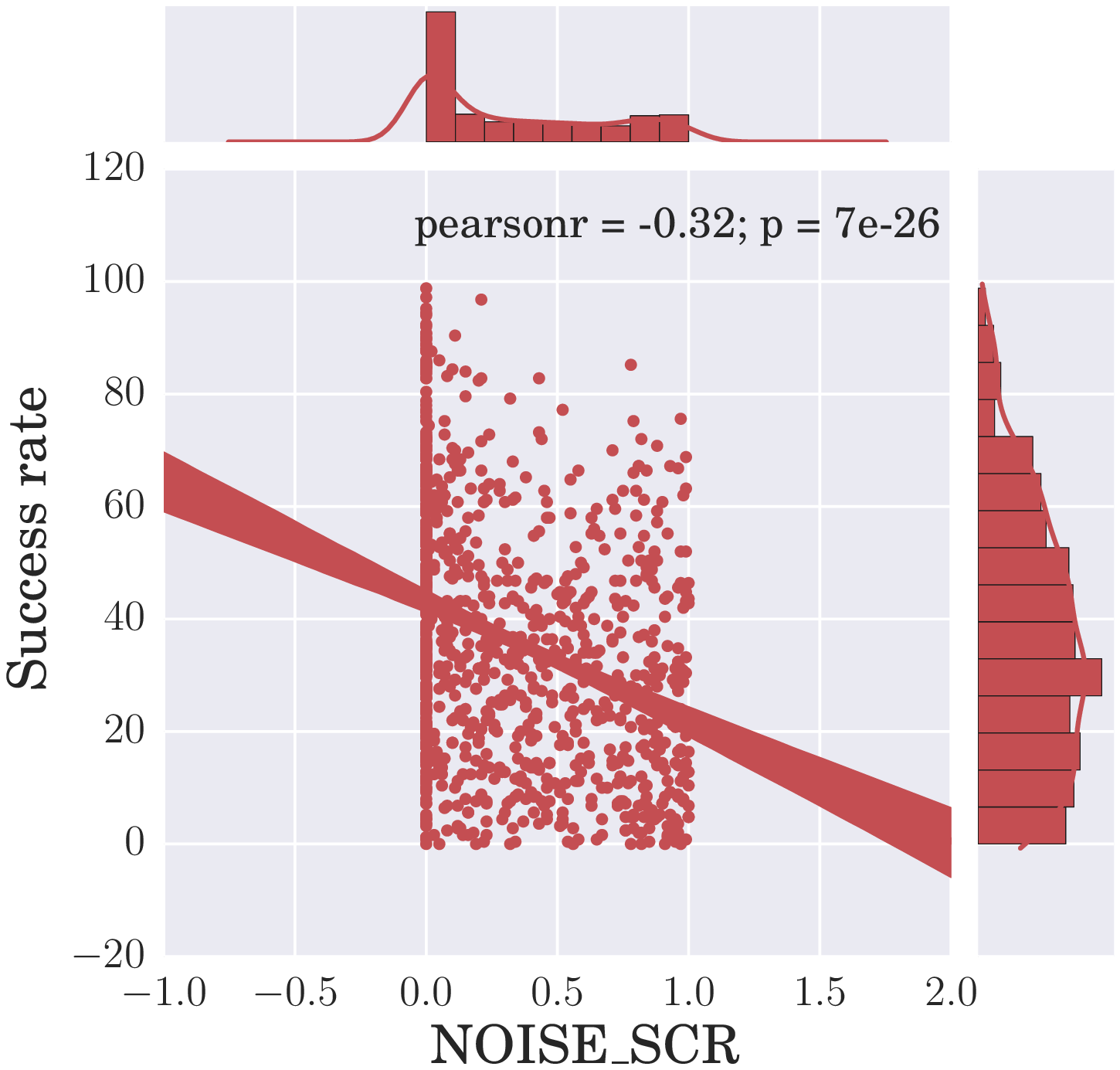}
 \includegraphics[width=5.5cm]{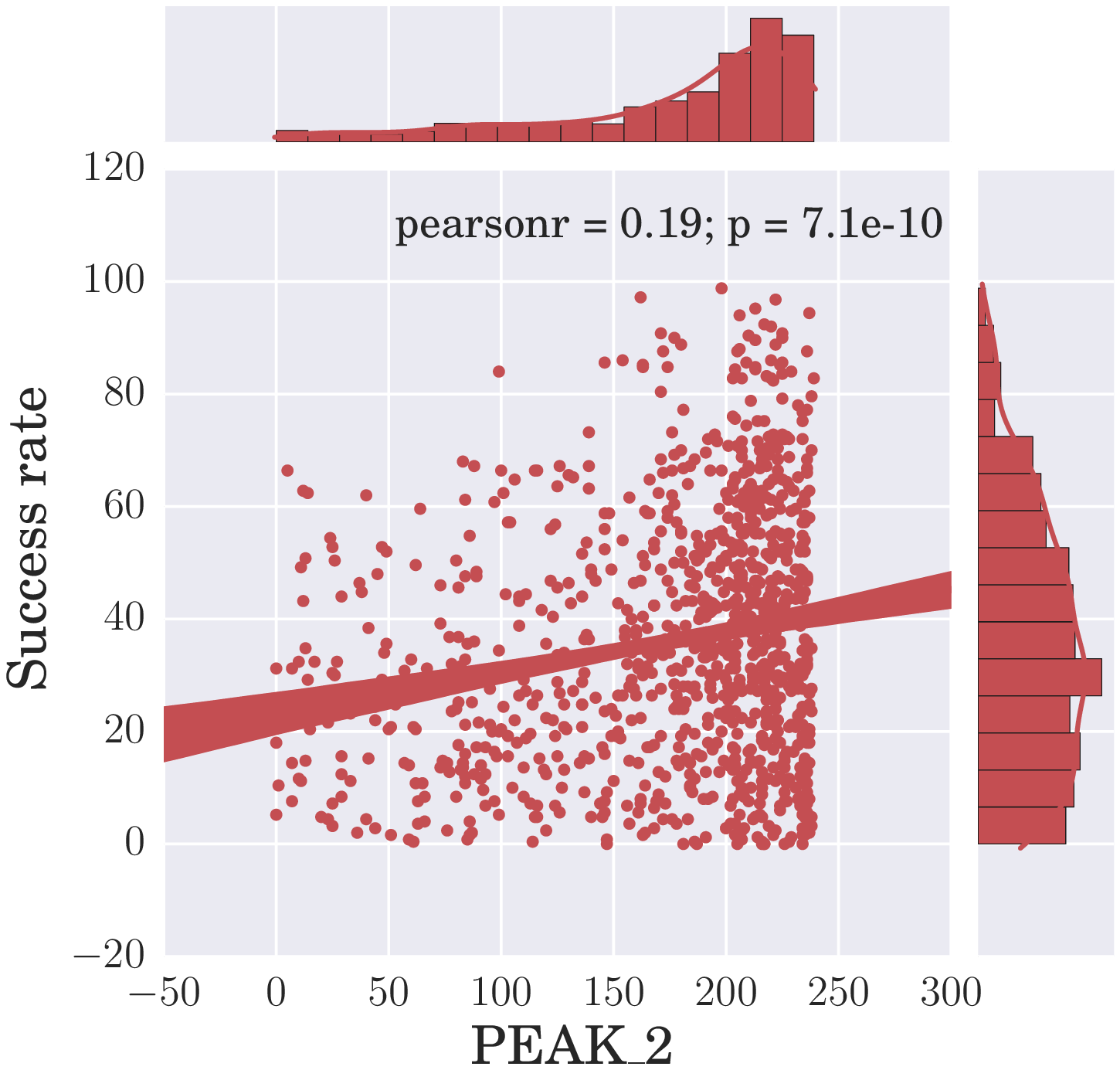}
\begin{pspicture}(0,0)(12.0,0.4)
    \rput(0.6,0.2){a)} \rput(6.1,0.2){b)} \rput(11.8,0.2) {c)}
\end{pspicture}
\caption{Pearson's correlation and linear approximation between the success rate of algorithm 1PCX-GA and three descriptors of the instances: a) $dJ+dh\_RES$; b) $NOISE\_SCR$; c) $PEAK\_2S$.}
\label{fig:PEARSON_ALG2}
\end{center}
\end{figure*}

 We computed the Pearson's correlations between the success rates of EAs and the descriptors of the instances. This information is shown in Table~\ref{tab:DES_SUCCRATE} where only a small number of correlations were not statistically significant using $\alpha=0.01$. Most of them involve algorithm FDA or descriptor  $HEIGHT\_2$. The strongest correlations were found for descriptors  dJ+dh\_RES and NOISE\_SCR. Figure~\ref{fig:PEARSON_ALG2} shows the patterns of the relationships between these descriptors and the success rate of 1PCX-GA for all instances. Also included in Figure~\ref{fig:PEARSON_ALG2} is the relationship for descriptor $PEAK\_2$, which showed significant correlations for all GAs but no significant correlations for any of the EDAs. The strong correlation between success rates and dJ+dh\_RES seems to suggest that the more the first excited states (low resilience), then there will be more local minima  where the algorithm will likely get trapped. Thus these instances are harder. Also the strong anticorrelation between success rates and  NOISE\_SCR suggests that instances with multiple peaks (high NOISE\_SCR)  are typically harder than the ones with one peak.

\subsection{Comparison with Simulated Annealing}

 As a final step we compared the results of the EAs to results achieved by SA. We ran two variants of SA proposed in \cite{Isakov_et_al:2015} and named $an\_ss\_rn\_fi$ and  $an\_ss\_ge\_nf\_bp$. Both variants are implemented for spin glasses with fixed number of neighbors but  $an\_ss\_ge\_nf\_bp$ is conceived to take advantage of the structure bipartite graphs. The two variants were run with the same parameters: Number SA sweeps $2000$, $\beta_0=0.1$, $\beta_1=3$, where $\beta=\frac{1}{T}$ is the inverse of the temperature, and $\beta_0$ and $\beta_1$ are the initial and final parameters of the linear schedule used for annealing. As in the case of the EAs these parameters are not expected to be optimal for all instances but we checked that increasing the number sweeps did not improve the results significantly.  For each Ising instance, $10000$ repetitions of the algorithm were executed and from these runs we computed the success rate.  

\begin{figure}[htbp]
  \begin{center}
   {\includegraphics[width=5.7cm]{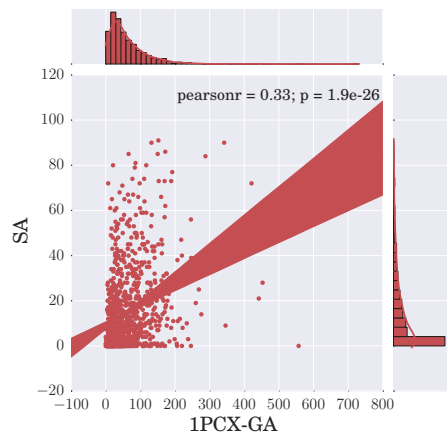}}
     \caption{Success rate of 1PCX-GA versus SA variant ($an\_ss\_ge\_nf\_bp$) designed for bipartite lattices \cite{Isakov_et_al:2015}. The number of successful runs is shown for each algorithm. $250$ runs were executed for 1PCX-GA and $10000$ for SA. }
    \label{fig:SRGASA}
    \end{center}
\end{figure}

 Results for  $an\_ss\_rn\_fi$ were very poor and therefore we present results here for $an\_ss\_ge\_nf\_bp$. Figure~\ref{fig:SRGASA} shows the correlation between the success rate of SA and 1PCX-GA. SA achieved a success rate above $0.25$ for only $51$ instances (versus $991$ for 1PCX-GA). However, it was able to find the optimum for all the instances at least once in $10000$ repetitions. SA is also orders of magnitude faster than EAs. It is very difficult to compare both optimizers since the way they apply partial evaluation and combine it with full evaluation of the solutions differ between the algorithms. Our results show that GAs are at least competitive with SA. More importantly, the analysis of Figure~\ref{fig:SRGASA} reveals that while there is a correlation between the hardness of the instances for both optimizers, 1PCX-GA exhibits a wider variability in its behavior and the sources of difficulty for both methods are not completely the same. Further experimental work is needed, using other problem benchmarks to assess the differences in the behavior between SA and 1PCX-GA. 

\section{Conclusions} \label{sec:CONCLU}

In this paper we have investigated for the first time the behavior of EAs on problems defined of the Chimera instances used by D-Wave architectures. We have shown that a simple GA with one-point crossover is able to solve $991$ of the $1000$ instances considered although the success rate of the algorithm depends on the instances. Our results show that EAs that use probabilistic modeling of the solutions dot not produce an improvement over methods that do not incorporate any type of modeling. To some extent this was an unexpected result because different variants of EDAs had shown good results for Ising problems defined on other topologies. Using problem information does not provide improvements in terms of success rate, although by restricting the number pair-wise dependencies to the edges of the Chimera graph important gains in terms of computational time were achieved by Tree-EDAr over Tree-EDA.

  We have also identified a number of instance descriptors which are correlated with the behavior of the algorithms. This could serve as a first step for a more complete characterization of the impact that certain features of the Ising instances have in the performance of EAs. An analysis of the impact of the same features for other optimizers could help to understand how different methods explore the space of solutions to identify the optimum.  

 Finally, although constructing instances whose pattern of interactions match that of the D-Wave architecture is a sensible way to investigate the performance of QA on the spin glass problem, it is not completely clear how the topological restrictions can affect the complexity of the instances and bias the behavior of other optimization algorithms applied to these instances. Our work could be used to advance the understanding of this and related issues. 

\section{Acknowledgments}
R. Santana's work has received support by the IT-609-13 program (Basque Government) and  TIN2013-41272P (Spanish Ministry of Science and Innovation). H.G.K~acknowledges support from the National Science Foundation (Grant No.~DMR-1151387). The work of H.G.K.~is supported in part by the Office of the Director of National Intelligence (ODNI), Intelligence Advanced Research Projects Activity (IARPA), via MIT Lincoln Laboratory Air Force Contract No.~FA8721-05-C-0002.  The views and conclusions contained herein are those of the authors and should not be interpreted as necessarily representing the official policies or endorsements, either expressed or implied, of ODNI, IARPA, or the U.S.~Government. The U.S.~Government is authorized to reproduce and distribute reprints for Governmental purpose notwithstanding any copyright annotation thereon.

\bibliographystyle{abbrv}  
\bibliography{Thesbib} 

\begin{thebibliography}{10}

\bibitem{Anderson_et_al:1991}
C.~A. Anderson, K.~F. Jones, and J.~Ryan.
\newblock A two-dimensional genetic algorithm for the {I}sing problem.
\newblock {\em Complex Systems}, 5(3):327--334, 1991.

\bibitem{Baluja:2006}
S.~Baluja.
\newblock Incorporating a priori knowledge in probabilistic-model based
  optimization.
\newblock In M.~Pelikan, K.~Sastry, and E.~Cant\'u-Paz, editors, {\em Scalable
  Optimization via Probabilistic Modeling: From Algorithms to Applications},
  Studies in Computational Intelligence, pages 205--222. Springer, 2006.

\bibitem{Baluja_and_Davies:1997r}
S.~Baluja and S.~Davies.
\newblock Using optimal dependency-trees for combinatorial optimization:
  Learning the structure of the search space.
\newblock In {\em Proceedings of the 14th International Conference on Machine
  Learning}, pages 30--38, 1997.

\bibitem{Brooke_et_al:1999}
J.~Brooke, D.~Bitko, R.~T. F., and G.~Aeppli.
\newblock Quantum annealing of a disordered magnet.
\newblock {\em Science}, 284(5415):779--781, 1999.

\bibitem{Chakrabarti_et_al:2005}
B.~K. Chakrabarti and A.~Das.
\newblock Transverse {I}sing model, glass and quantum annealing.
\newblock In {\em Quantum Annealing and Other Optimization Methods}, pages
  1--36. Springer, 2005.

\bibitem{Dahl:2013}
E.~D. Dahl.
\newblock Programming with {D-Wave}: {M}ap coloring problem.
\newblock White paper, DWave. The Quantum Computing Company, 2013.

\bibitem{Dash:2013}
S.~Dash.
\newblock A note on {QUBO} instances defined on {Chimera} graphs.
\newblock {\em arXiv preprint arXiv:1306.1202}, 2013.

\bibitem{Denchev_et_al:2015}
V.~S. Denchev, S.~Boixo, S.~V. Isakov, N.~Ding, R.~Babbush, V.~Smelyanskiy,
  J.~Martinis, and H.~Neven.
\newblock What is the computational value of finite range tunneling?
\newblock {\em arXiv preprint arXiv:1512.02206}, 2015.

\bibitem{Echegoyen_et_al:2010}
C.~Echegoyen, A.~Mendiburu, R.~Santana, and J.~A. Lozano.
\newblock Estimation of {B}ayesian networks algorithms in a class of complex
  networks.
\newblock In {\em Proceedings of the 2010 Congress on Evolutionary Computation
  CEC-2010}, Barcelone, Spain, 2010. IEEE Press.

\bibitem{Hamze_and_deFreitas:2004}
F.~Hamze and N.~de~Freitas.
\newblock From fields to trees.
\newblock In {\em Uncertainty in Artificial Intelligence (UAI)}, pages
  243--250, Arlington, Virginia, 2004. AUAI Press.

\bibitem{Harik_et_al:2006}
G.~R. Harik, F.~G. Lobo, and K.~Sastry.
\newblock Linkage learning via probabilistic modeling in the {ECGA}.
\newblock In M.~Pelikan, K.~Sastry, and E.~Cant\'u-Paz, editors, {\em Scalable
  Optimization via Probabilistic Modeling: From Algorithms to Applications},
  Studies in Computational Intelligence, pages 39--62. Springer, 2006.

\bibitem{Isakov_et_al:2015}
S.~V. Isakov, I.~N. Zintchenko, T.~F. R{\o}nnow, and M.~Troyer.
\newblock Optimised simulated annealing for {I}sing spin glasses.
\newblock {\em Computer Physics Communications}, 192:265--271, 2015.

\bibitem{Kadowaki_and_Nishimori:1998}
T.~Kadowaki and H.~Nishimori.
\newblock Quantum annealing in the transverse {I}sing model.
\newblock {\em Physical Review E}, 58(5):5355, 1998.

\bibitem{Katzgraber_et_al:2015}
H.~G. Katzgraber, F.~Hamze, Z.~Zhu, A.~J. Ochoa, and H.~Munoz-Bauza.
\newblock Seeking quantum speedup through spin glasses: The good, the bad, and
  the ugly*.
\newblock {\em Phys. Rev. X}, 5:031026, Sep 2015.

\bibitem{Maksymowicz_et_al:1994}
A.~Maksymowicz, J.~Galletly, M.~Magdon, and I.~Maksymowicz.
\newblock Genetic algorithm approach for {I}sing model.
\newblock {\em Journal of magnetism and magnetic materials}, 133(1):40--41,
  1994.

\bibitem{Mandra_et_al:2016}
S.~Mandr{\`a}, Z.~Zhu, W.~Wang, A.~Perdomo-Ortiz, and H.~G. Katzgraber.
\newblock Strengths and weaknesses of weak-strong cluster problems: A detailed
  overview of state-of-the-art classical heuristics vs quantum approaches.
\newblock {\em arXiv preprint arXiv:1604.01746}, 2016.

\bibitem{Mcgeoch_and_Wang:2013}
C.~C. McGeoch and C.~Wang.
\newblock Experimental evaluation of an adiabiatic quantum system for
  combinatorial optimization.
\newblock In {\em Proceedings of the ACM International Conference on Computing
  Frontiers}, page~23. ACM, 2013.

\bibitem{Muhlenbein_et_al:1999}
H.~M{\"{u}}hlenbein, T.~Mahnig, and A.~Ochoa.
\newblock Schemata, distributions and graphical models in evolutionary
  optimization.
\newblock {\em Journal of Heuristics}, 5(2):213--247, 1999.

\bibitem{Pelikan_and_Hartmann:2006}
M.~Pelikan and A.~K. Hartmann.
\newblock Searching for ground states of {I}sing spin glasses with hierarchical
  {BOA} and cluster exact approximation.
\newblock In M.~Pelikan, K.~Sastry, and E.~Cant\'u-Paz, editors, {\em Scalable
  Optimization via Probabilistic Modeling: From Algorithms to Applications},
  Studies in Computational Intelligence, pages 333--349. Springer, 2006.

\bibitem{Pelikan_and_Katzgraber:2009}
M.~Pelikan and H.~G. Katzgraber.
\newblock Analysis of evolutionary algorithms on the one-dimensional spin glass
  with power-law interactions.
\newblock In {\em Proceedings of the 11th Annual conference on Genetic and
  evolutionary computation}, pages 843--850. ACM, 2009.

\bibitem{Prugel_and_Shapiro:1997}
A.~Pr{\"u}gel-Bennett and J.~L. Shapiro.
\newblock The dynamics of a genetic algorithm for simple random {I}sing
  systems.
\newblock {\em Physica D: Nonlinear Phenomena}, 104(1):75--114, 1997.

\bibitem{Santana:2005}
R.~Santana.
\newblock {E}stimation of distribution algorithms with {K}ikuchi
  approximations.
\newblock {\em Evolutionary Computation}, 13(1):67--97, 2005.

\bibitem{Santoro_et_al:2002}
G.~E. Santoro, R.~Marto{\v{n}}{\'a}k, E.~Tosatti, and R.~Car.
\newblock Theory of quantum annealing of an {I}sing spin glass.
\newblock {\em Science}, 295(5564):2427--2430, 2002.

\bibitem{Selby:2014}
A.~Selby.
\newblock Efficient subgraph-based sampling of {I}sing-type models with
  frustration.
\newblock {\em arXiv preprint arXiv:1409.3934}, 2014.

\bibitem{Shakya_et_al:2006}
S.~Shakya, J.~McCall, and D.~Brown.
\newblock Solving the {I}sing spin glass problem using a bivariate {EDA} based
  on {M}arkov random fields.
\newblock In {\em Proceedings of the IEEE Congress on Evolutionary Computation.
  CEC-2006}, pages 908--915. IEEE, 2006.

\bibitem{Shakya_and_Santana:2012}
S.~Shakya and R.~Santana, editors.
\newblock {\em Markov Networks in Evolutionary Computation}.
\newblock Springer, 2012.

\bibitem{Zhu_et_al:2015a}
Z.~Zhu, A.~J. Ochoa, and H.~G. Katzgraber.
\newblock Efficient cluster algorithm for spin glasses in any space dimension.
\newblock {\em Phys. Rev. Lett.}, 115:077201, 2015.

\bibitem{Zhu_et_al:2015}
Z.~Zhu, A.~J. Ochoa, S.~Schnabel, F.~Hamze, and H.~G. Katzgraber.
\newblock Best-case performance of quantum annealers on native spin-glass
  benchmarks: How chaos can affect success probabilities.
\newblock {\em arXiv preprint arXiv:1505.02278}, 2015.

\bibitem{Zintchenko_et_al:2015}
I.~Zintchenko, M.~B. Hastings, and M.~Troyer.
\newblock From local to global ground states in {I}sing spin glasses.
\newblock {\em Physical Review B}, 91(2):024201, 2015.

\end{thebibliography}

\end{document}